\documentclass[twocolumn,aps,prd,amsmath,amssymb]{revtex4-1}
\usepackage[latin9]{inputenc}
\usepackage{amstext}
\usepackage{amssymb}
\usepackage{graphicx}
\usepackage{esint}

\makeatletter
\usepackage{amssymb}
\usepackage{mathbbold}
\usepackage{amsfonts}
\usepackage{epsfig}
\usepackage{bm}\usepackage{dcolumn}\usepackage{color}
\usepackage{overpic}
\usepackage[bottom]{footmisc}

\makeatother

\begin{document}
	
\title{Chiral vortices and pseudoscalar condensation due to rotation}

\author{Lingxiao Wang$^{1}$}

\author{Yin Jiang$^{2}$}

\author{Lianyi He$^{1}$}

\author{Pengfei Zhuang$^{1}$}

\affiliation{$^{1}$Department of Physics, Tsinghua University, Beijing 100084, China\\
$^{2}$Department of Physics, Beihang University, Beijing 100191, China}

\date{\today}

\begin{abstract}
We investigate the influence of rotation on the dynamical chiral symmetry breaking in strongly interacting matter.  We develop a self-consistent Bogoliubov-de Gennes-like theoretical framework to study the
inhomogeneous chiral condensate and the possible chiral vortex state in rotating finite-size matter in four-fermion interacting theories.  We show that for sufficiently 
rapid rotation in $2+1$ dimensions, the ground state can be a chiral vortex state, a type of topological defect in analogy to superfluids and superconductors.  The vortex state exhibits pion condensation,
providing a new mechanism to realize pseudoscalar condensation in strongly interacting matter. 
\end{abstract}


\maketitle

%%%%%%%%%%%%%%%%%%%%%%%%%%%%%%%%%%%%%
\section{Introduction}\label{s1}
%%%%%%%%%%%%%%%%%%%%%%%%%%%%%%%%%%%%%

A good knowledge of Quantum Chromodynamics (QCD) in extreme conditions is crucial for us to understand a wide range of physical phenomena. The QCD phase diagram at finite
temperature and density, related to the early Universe and the compact stars, has been comprehensively investigated~\cite{Review01,Review02,Review03,Review04,Review05}. In relativistic heavy ion collisions,  it is generally believed that strong electromagnetic field and vorticity can be generated in non-central collisions~\cite{EM01,EM02,EM03,EM04,EM05,EM06,EM07,Vorticity01,Vorticity02,Vorticity03,Vorticity04,Vorticity05,Vorticity06,Vorticity07}.  Recently, a direct experimental evidence for strong vorticity, the global 
$\Lambda$ hyperon polarization~\cite{Vorticity01,Vorticity02}, has been observed in high energy nuclear collision~\cite{Vorticity07}.  In the high-temperature quark-gluon plasma phase,  the QCD matter shows intriguing features at strong magnetic field and/or vorticity, such as the chiral magnetic effect and the chiral vortical effect, induced by the quantum anomaly of chiral fermions~\cite{Huang2016,Kharzeev2015}. On the other hand, understanding the phase structure of strongly interacting matter at finite electromagnetic field and vorticity also becomes important.

In this work, we focus on the chiral phase structure at finite rotation, which can be in principle studied by using lattice QCD~\cite{Lattice-Rotation}. It is widely accepted that the chiral symmetry breaking and restoration can be well described in some low-energy effective models~\cite{NJL-Review01,NJL-Review02,NJL-Review03}. Here we study the chiral phase structure of strongly interacting matter under rotation using a four-fermion interacting model, known as the Nambu-Jona--Lasinio (NJL) model~\cite{NJL}.  In rotating relativistic matter, the finite size and boundary effects cannot be ignored~\cite{NJL-Rotation01,NJL-Rotation02,NJL-Rotation03}, and the chiral condensate is generically inhomogeneous.  In the previous works,  the local density approximation (LDA) is employed to evaluate the quasiparticle spectrum and the free energy \cite{NJL-Rotation01,NJL-Rotation02,NJL-Rotation03,NJL-Rotation04,NJL-Rotation05,NJL-Rotation06,NJL-Rotation07,NJL-Rotation08}. In LDA, the chiral condensate is treated as locally homogeneous, so that the single-particle Dirac equation can be solved analytically.  Its validity thus depends on the system size and 
it may become invalid for small system sizes. Furthermore,  such an approximation excludes the possibility of some exotic phases, such as the quantized vortex state~\cite{Vortex01,Vortex02,Vortex03,Vortex04,Vortex05,Vortex06,Vortex07,Vortex08}, a type of topological defect associated with the spontaneous symmetry breaking.

One purpose of this work is to develop a self-consistent theoretical framework to study the inhomogeneous chiral condensate in rotating finite-size matter in four-fermion interacting theories. Keeping in mind that the
dynamical chiral symmetry breaking in four-fermion interacting models is analogous to the BCS superconductivity,  we use a Bogoliubov-de Gennes (BdG) approach~\cite{BdG} for inhomogeneous chiral condensate in a finite-size rotating system. In such an approach, the single-particle spectrum and the inhomogeneous chiral condensate are self-consistently determined by solving the BdG equation numerically.

On the other hand, rotating quantum matter can exhibit interesting physical phenomena, such as quantized vortices and vortex lattices, which has been observed in rapidly rotating Bose-Einstein condensates and fermionic superfluids~\cite{Vortex-BEC01,Vortex-BEC02,Vortex-BEC03,Vortex-Fermion}.  A quantized vortex is a type of topological defect, which can appear in systems with spontaneous breaking of continuous symmetry.  The simplest scenario is the spontaneous breaking of a U(1) symmetry, such as in Bose-Einstein condensates and fermionic superfluids. In this case, the circulation around the vortex axis is provided by the phase of the complex order parameter. 

The self-consistent BdG theory enables us to explore the vortex solution of the chiral symmetry breaking in rotating strongly interacting matter. In this work, we study the NJL model with the simplest U(1) chiral symmetry. 
The vortex solution in a rotating $(2+1)$-dimensional NJL system is obtained by solving the BdG equation. We find that for a finite-size system, the vortex state can be the ground state when the angular velocity becomes larger than a critical value. Moreover, in the vortex state, nonzero expectation value of the pion field is generated. Therefore, rotation provides a new route to generate pseudoscalar condensation in strongly interacting matter.

The paper is organized as follows.  In Sec. \ref{s2}, we set up the general BdG theoretical framework for a four-fermion interaction model with U(1) chiral symmetry.  In Sec. \ref{s3}, we present the computational details for a $(2+1)$-dimensional system. We discuss the numerical results in Sec. \ref{s4} and summarize in Sec. \ref{s5}. The natural units $c=\hbar=k_{\rm B}=1$ will be used throughout. 

%%%%%%%%%%%%%%%%%%%%%%%%%%%%%%%%%%%%%
\section{Bogoliubov-de Gennes theory}\label{s2}
%%%%%%%%%%%%%%%%%%%%%%%%%%%%%%%%%%%%%

We study a four-fermion interaction model with a U(1) chiral symmetry. The Lagrangian density is given by
\begin{equation}
\mathcal{L}=\bar{\psi}i\gamma^\mu\partial_\mu\psi+\frac{G}{2N}\left[(\bar{\psi}\psi)^2+(\bar{\psi}i\gamma_5\psi)^2\right] 
\end{equation}
where the fermion field $ \psi$ is a four-component spinor, $N$ is the number of flavor, and $ G $ is the coupling constant.  We consider a finite-size system with a constant angular velocity along the $z$ direction, 
$\mbox{\boldmath{$\omega$}}=\omega\hat{\textbf{z}}$. The theory thus can be considered in either 3+1 or 2+1 dimensions. 

We evaluate the partition function ${\cal Z}$ in the rotating frame. In the imaginary time formalism, it can be expressed as
\begin{equation}
\mathcal{Z}=\int[d\psi^\dagger][d\psi]e^{-{\cal S}[\psi^\dagger,\psi]},
\end{equation}
with the action 
\begin{equation}
{\cal S}[\psi^\dagger,\psi]=\int dx\sqrt{-\det g_{\mu\nu}}\left[\psi^\dagger\partial_\tau\psi+\mathcal{H}(\psi^\dagger,\psi)\right].
\end{equation}
Here $\int dx=\int_0^\beta d\tau\int d{\bf r}$ with $\tau$ being the imaginary time and $\beta$ being the inverse of the temperature $T$, and $g_{\mu\nu}$ is the space-time metric of the rotating frame~\cite{Rotating-Frame01,Rotating-Frame02}. The Hamiltonian density in the rotating frame is given by
\begin{equation}
\mathcal{H}=\psi^\dagger\hat{K}_0\psi-\frac{G}{2N}\left[(\bar{\psi}\psi)^2+(\bar{\psi}i\gamma_5\psi)^2\right],
\end{equation}
where 
\begin{equation}
\hat{K}_0=-i\gamma^0\mbox{\boldmath{$\gamma$}}\cdot\mbox{\boldmath{$\nabla$}}-\mbox{\boldmath{$\omega$}}\cdot\hat{\bf J}.
\end{equation}
Here the the single-particle's angular momentum 
is given by $\hat{\bf J}=-i{\bf r}\times\mbox{\boldmath{$\nabla$}}+\frac{1}{2}\mbox{\boldmath{$\Sigma$}}$, with $\mbox{\boldmath{$\Sigma$}}$ being the spin operator.

We evaluate the partition function to the leading-order in the $1/N$ expansion, which amounts to the mean-field approximation. This approximation becomes accurate in the large $N$ ($N\rightarrow\infty$) limit. We thus define the following scalar (sigma) and pseudoscalar (pion) condensates,
\begin{equation}
\sigma({\bf r})=-\frac{G}{N}\langle\bar{\psi}\psi\rangle, \ \ \ \ \pi({\bf r})=-\frac{G}{N}\langle\bar{\psi}i\gamma_5\psi\rangle.
\end{equation}
Note that the above expectation values can be in principle spatially inhomogeneous. The Hamiltonian density in the mean-field approximation reads
\begin{equation}
\mathcal{H}_{\rm MF}=\psi^\dagger \hat{K}[\sigma, \pi]\psi+\frac{N}{2G}\left(\sigma^2+\pi^2\right),
\end{equation}
where $ \hat{K}=\hat{K}_0+\gamma^0(\sigma+i\gamma_5\pi) $. Furthermore, because of the U(1) chiral symmetry, the sigma and pion condensates can be combined into a complex order parameter
\begin{equation}
\Delta({\bf r})=\sigma({\bf r})+i \pi({\bf r})=M({\bf r})e^{i \phi({\bf r})},
\end{equation}
where the modulus $M({\bf r})$ and the phase $\phi({\bf r})$ are set to be real. For an infinite uniform system, spontaneous chiral symmetry breaking is characterized by a uniform solution $M\neq0$, and the phase
can be chosen arbitrarily.  Physically we chose $\phi=0$, indicating a vanishing pseudoscalar condensate $\pi=0$.

For a given profile of $ \Delta({\bf r})$, the partition function in mean-field approximation can be evaluated once the eigenvalue problem of the operator
$ \hat{K} $ is solved. The eigenvalue equation can be expressed as 
\begin{equation}
\hat{K}\Psi_n({\bf r})=\varepsilon_n\Psi_n({\bf r}),
\label{eigenvalue-equation}
\end{equation}
where $n$ is a complete set of quantum numbers. Using the transform $ \psi(\tau,{\bf r})=\sum_{n,k}c_{nk}e^{-i\omega_k\tau}\Psi_n({\bf r})$, the action in the mean-field approximation can be evaluated as 
\begin{equation}
{\cal S}_{\rm MF}=\frac{\beta N}{2G}\int d{\bf r}|\Delta({\bf r})|^2-\beta\sum_{n,k}c^*_{nk}(i\omega_k-\varepsilon_n)c_{nk}^{\phantom{\dag}},
\end{equation}
where $\omega_k=(2k+1)\pi T$ ($k\in\mathbb{Z}$) is the fermion Matsubara frequency.
The free energy $F$ can be evaluated by completing the functional integral over $c_{nk}^*$ and $c_{nk}^{\phantom{\dag}}$. We obtain
\begin{equation}
\frac{F}{N}=\frac{1}{2G}\int d{\bf r}|\Delta({\bf r})|^2-\sum_n \left[ \frac{\varepsilon_n}{2}+\frac{1}{\beta}\ln\left(1+e^{-\beta \varepsilon_n}\right)\right].
\label{potential}
\end{equation}
The eigenvalue equation (\ref{eigenvalue-equation}) is nothing but the Bogoliubovde Gennes equation for an inhomogeneous condensate $ \Delta({\bf r}) $. The inhomogeneous profile $ \Delta({\bf r}) $ should be self-consistently determined by the variational condition
\begin{equation}
\frac{\delta F[\Delta({\bf r})]}{\delta \Delta({\bf r})}=0.
\label{variational-condition}
\end{equation}
For the homogeneous case, this equation gives nothing but the so-called gap equation. Equations (\ref{eigenvalue-equation}) and (\ref{variational-condition}), together with some proper boundary condition, constitute a type of Bogoliubov-de Gennes theory for the present rotating finite-size system.

We are interested in a general solution with an arbitrary circulation number $\kappa$,  which is related to the phase of the condensate, $\oint d{\bf l}\cdot\mbox{\boldmath{$\nabla$}}\phi=2\pi\kappa $. 
Working in cylindrical coordinates $ {\bf r}=(\rho,\theta,z) $, we look for the solution of the following form,
\begin{equation}
\Delta({\bf r})=M(\rho)e^{i\kappa \theta}, \ \ \ \ \kappa\in\mathbb{Z}.
\label{vortex-form}
\end{equation}
The corresponding sigma and pion condensates read
\begin{equation}
\sigma({\bf r})=M(\rho)\cos(\kappa\theta), \ \ \ \ \pi({\bf r})=M(\rho)\sin(\kappa\theta).
\end{equation}
In the previous works, the trivial case $\kappa=0$ was considered. The solution with a nonvanishing circulation, $\kappa\neq0$, corresponds to the quantized
vortex state. In this state, an angular stripe-like pion condensate is generated.

%%%%%%%%%%%%%%%%%%%%%%%%%%%%%%%%%%%%%
\section{ (2+1)-dimensional system}\label{s3}
%%%%%%%%%%%%%%%%%%%%%%%%%%%%%%%%%%%%%

The BdG theoretical framework presented in Sec. \ref{s2} is applicable for both $3+1$ and $2+1$ dimensions. However, it is known that in $3+1$ dimensions, the NJL-type four-fermions interaction model
is not renormalizable.  In this case, we need to introduce a specific regularization scheme and this brings scheme and parameter dependence (see Appendix A). Moreover, in the cylindrical system we considered, the longitudinal $z$ degree of
freedom also leads to a huge computational cost. On the other hand, we expect that the rotation effect on the chiral condensate in 2+1 dimensions is similar to that in 3+1 dimensions. Therefore, we will consider the $(2+1)$-dimensional system in this work.

The advantage of the $(2+1)$-dimensional system is that the four-fermion interaction model can be renormalized to arbitrary order in $1/N$ \cite{NJL-3D}. Thus the artificial effects, like the regularization scheme dependence and the model parameter dependence can be completely avoided in 2+1 dimensions. The model is renormalized in vacuum. It is obvious that rotation, finite size, and finite temperature effects will not cause new ultraviolet divergence.  In the uniform vacuum state, we choose $\sigma=M$ and $\pi=0$. At the leading order in $1/N$, the effective potential $V$ can be evaluated as
\begin{equation}
\frac{V}{N}=\frac{M^2}{2G}-2\int\frac{d^2{\bf k}}{(2\pi)^2}\sqrt{{\bf k}^2+M^2}.
\end{equation}
The integral over the momentum ${\bf k}$ is divergent and we introduce a cutoff $\Lambda$. For large cutoff ($\Lambda\rightarrow\infty$), the effective potential reads
\begin{equation}
\frac{V}{N}=\frac{M^2}{2}\left(\frac{1}{G}-\frac{\Lambda}{\pi}\right)+\frac{M^3}{3\pi},
\end{equation}
where we have dropped a vacuum term independent of $M$. Therefore, at the leading order in $1/N$, only the coupling constant $G$ needs renormalization. The bare coupling constant $ G(\Lambda) $ should be fine tuned 
such that~\cite{NJL-3D}
\begin{equation}
\frac{1}{G(\Lambda)}-\frac{1}{G_c}=-\frac{M_0}{\pi}{\rm sgn}(G-G_c),
\label{renormalization}
\end{equation}
where the critical coupling is given by $ G_c=\pi/\Lambda$. The emergent finite quantity $M_0>0$ serves as the only mass scale in the theory.  In this work we focus on the case $G>G_c$. In this case, spontaneous chiral symmetry breaking occurs in vacuum and the effective fermion mass is given by $ M_*=M_0 $.

The BdG equation can be explicitly expressed as   
\begin{equation}
\left( \begin{array}{cc}
\hat{K}_{11} & \hat{K}_{12} \\ 
\hat{K}_{21}& \hat{K}_{22}
\end{array} \right)
\left(\begin{array}{c}
		u_{n}({\bf r})\\ 
		v_{n}({\bf r}) \end{array}\right)
=\varepsilon_n\left(\begin{array}{c}
		u_{n}({\bf r})\\ 
		v_{n}({\bf r}) \end{array}\right).
\end{equation}
Starting from the usual gamma matrices for four-component spinor in 2+1 dimensions \cite{Gamma-3D},  we arrive at a chiral-like representation after a unitary transform. In this representation, the gamma matrices are given by
\begin{eqnarray}
	&&\gamma^0=\left( \begin{array}{cc}
		0 & \sigma_3 \\ 
		\sigma_3 & 0
	\end{array} \right), \ \ \ \ \ 
	 \gamma^1=\left(\begin{array}{cc}
		0 & -i \sigma_1 \\ 
		-i\sigma_1 & 0
	\end{array}  \right), \nonumber\\
	&&\gamma^2=\left( \begin{array}{cc}
		0 & i\sigma_2 \\ 
		i\sigma_2 & 0
	\end{array} \right) , \ \ \ 
	\gamma^5=\left( \begin{array}{cc}
		-I & 0 \\ 
		0 & I
	\end{array} \right). \nonumber
\end{eqnarray}
Here $ I $ is the $ 2\times2 $ identity matrix and $\sigma_i$ ($i=1,2,3$)
are the Pauli matrices. The blocks of the $\hat{K}$ operator are
then given by 
\begin{eqnarray}
&&\hat{K}_{11}=\hat{K}_{22}=-i(\sigma_2 \partial_x+\sigma_1\partial_y)-\omega (\hat{l}_z+\sigma_3/2),\nonumber\\
&&\hat{K}_{12}=\sigma_3\Delta({\bf r}),\ \ \ \ \ \ \  \hat{K}_{21}=\sigma_3\Delta^*({\bf r})
\end{eqnarray}
The advantage of this chiral representation is that the sigma and pion condensates are explicitly combined into a complex condensate $\Delta({\bf r})$, which appears in the off-diagonal blocks, in analogy to the BCS theory of superconductivity.  

We consider a circular box of radius $R$. Because of the rotational symmetry of the solution (\ref{vortex-form}), the wave functions for an arbitrary circulation $\kappa$ can be expressed as
\begin{eqnarray}
&&u_n({\bf r})=\sum_l\frac{e^{i l\theta}}{\sqrt{2\pi}}\left(\begin{array}{c}
		u_{nl}^\uparrow(\rho)e^{i \theta}\\ 
		u_{nl}^{\downarrow}(\rho) \end{array}\right),\nonumber\\
&&v_n({\bf r})=e^{-i\kappa\theta}\sum_l\frac{e^{i l\theta}}{\sqrt{2\pi}}\left(\begin{array}{c}
		v_{nl}^{\uparrow}(\rho) e^{i\theta}\\ 
		v_{nl}^{\downarrow}(\rho)\end{array}\right),
\end{eqnarray}
where $l$ denotes the angular quantum number. The BdG equation thus decouples into different $l$ sectors. To solve the BdG equation, we need a proper boundary condition at $\rho=R$.  To this end, we consider the current conservation in this finite size system. In the rotating frame the vector current conservation law reads
\begin{equation}
\nabla_\mu j^\mu=\frac{1}{\sqrt{|g|}}\partial_\mu\left(\sqrt{|g|}j^\mu\right)=0,
\end{equation}
where $\nabla_\mu$ is the covariant derivative, $j^\mu=\bar{\psi}\gamma^\mu\psi$, and $|g|=-\det g_{\mu\nu}$. To keep the total charge constant in the circle, we must impose a condition of no incoming flux at the spatial boundary. In the polar coordinates, this condition can be expressed as 
\begin{equation}\label{current}
R\int_{0}^{2\pi}d\theta\bar{\psi}\gamma^\rho\psi\Big|_{\rho=R}=0,
\end{equation}
where $ \gamma^\rho\equiv\gamma^1\cos\theta+\gamma^2\sin\theta $~\cite{NJL-Rotation01}.  It has been shown that this condition guarantees that the Hamiltonian of the system is self-adjoint and provides necessary and sufficient conditions for a set of boundary conditions to yield a consistent quantization.  However,  the boundary condition cannot be uniquely determined by Eq. (\ref{current}).  Different boundary conditions which satisfy Eq. (\ref{current}) has been comprehensively studied for noninteracting rotating fermions~\cite{Rotating-Frame02} and interacting rotating fermions~\cite{NJL-Rotation02}.

%%%%%%%%%%%%%%%%%%%%%%%%%%%%%%%%%%%%%%%%%%%%%%%%%%%%%%%%%%%%%%%%%%%%%%%
\begin{figure}
\centering{}\includegraphics[width=0.48\textwidth]{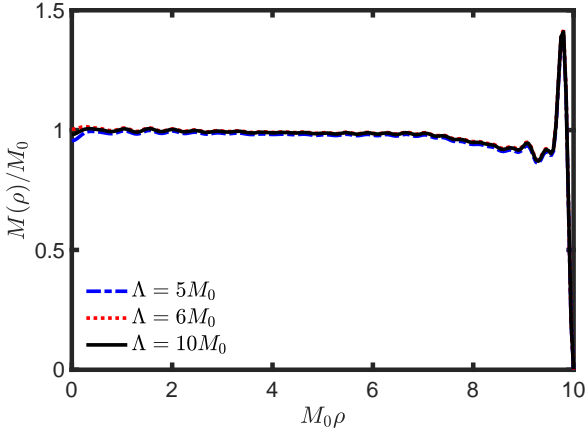} 
\caption{ Convergence of the calculations in $2+1$ dimensions.  This figure shows the profile $M(\rho)$ for various values of the cutoff $\Lambda$ for a system size $M_0R=10$ at zero temperature and vanishing rotation.
\label{fig1}}
\end{figure}

%%%%%%%%%%%%%%%%%%%%%%%%%%%%%%%%%%%%%%%%%%%%%%%%%%%%%%%%%%%%%%%%%%%%%%%% 

In this work, we emply a simple version of the boundary conditions which was previously used by Ebihara, Fukushima, and Mameda~\cite{NJL-Rotation01}. In this scheme, we define the orthonormal basis function 
\begin{equation}\label{basis}
\phi_{j,l}(\rho)=\frac{\sqrt{2}J_l(\alpha_{j,l}\rho/R)}{RJ_{l+1}(\alpha_{j,l})} ,
\end{equation}
with $ \alpha_{j,l} $ being the $ j$-th zero of the Bessel function $ J_l(x) $.  The radial parts of the wave functions are expanded as
\begin{eqnarray}
&&u_{nl}^\uparrow(\rho)=\sum_j c_{nj}^\uparrow\phi_{j,l+1}(\rho),\nonumber\\ 
&&u_{nl}^\downarrow(\rho)=\sum_j c_{nj}^\downarrow\phi_{j,l}(\rho),\nonumber\\
&&v_{nl}^\uparrow(\rho)=\sum_j d_{nj}^\uparrow\phi_{j,l-\kappa+1}(\rho),\nonumber\\  
&&v_{nl}^\downarrow(\rho)=\sum_j d_{nj}^\downarrow\phi_{j,l-\kappa}(\rho).
\end{eqnarray}
Since $\phi_{j,l}(R)=0$, the condition (\ref{current}) is satisfied. For a given $l$, the BdG equation thus reduces to a matrix form
\begin{widetext}
\begin{equation}
\sum_{j'}\left( \begin{array}{cccc}
		-K_{l+1}^{jj'} & S_{l}^{jj'} & \Delta_{l+1}^{jj'} & 0 \\ 
		S_{l}^{j'j} & K_{-l}^{jj'} & 0 & -\Delta_l^{jj'} \\ 
		\Delta_{l+1}^{j'j} & 0 & -K_{l-\kappa+1}^{jj'} & S_{l-\kappa}^{jj'} \\ 
		0 & -\Delta_l^{j'j} & S_{l-\kappa}^{j'j} & K_{-(l-\kappa)}^{jj'}
	\end{array} \right) \left(\begin{array}{c}
		c_{nj'}^{\uparrow} \\ 
		c_{nj'}^{\downarrow} \\ 
		d_{nj'}^{\uparrow} \\ 
		d_{nj'}^{\downarrow}
	\end{array}  \right)=\varepsilon_{nl}\left(\begin{array}{c}
		c_{nj}^{\uparrow} \\ 
		c_{nj}^{\downarrow} \\ 
		d_{nj}^{\uparrow} \\ 
		d_{nj}^{\downarrow}
	\end{array}  \right),
\label{BdG-Matrix}
\end{equation}
\end{widetext}
where the elements are given by 
\begin{eqnarray}
&&K_l^{jj'}=\omega(l-1/2)\delta_{jj'},\nonumber\\
&&S_l^{jj'}=\int d\rho \phi_{j,l+1}(\rho)\left(l-\rho \frac{\partial}{\partial\rho}\right)\phi_{j',l}(\rho),\nonumber\\
&&\Delta_l^{jj'}=\int \rho d\rho M(\rho)\phi_{j,l}(\rho)\phi_{j',l-\kappa}(\rho)
\end{eqnarray}

While different $l$ sectors are decoupled in (\ref{BdG-Matrix}), they are coupled through the variational equation (\ref{variational-condition}).
The variational equation can be explicitly given by
\begin{equation}
\frac{M(\rho)}{G(\Lambda)}=\sum_{n,l} \left[u^{\uparrow}_{nl}v^{\uparrow}_{nl}(\rho)-u^{\downarrow}_{nl}v^{\downarrow}_{nl}(\rho)\right] (1-2n_{\rm F}(\varepsilon_{nl})),
\label{2dgap}
\end{equation}
where $n_{\rm F}(\varepsilon)=1/(e^{\beta\varepsilon}+1)$ is the Fermi-Dirac distribution. Since the model is renormalizable, we expect that the dependence on the cutoff $\Lambda$ disappears once we set 
$\Lambda\rightarrow\infty$.  In practice, we impose a high-energy cutoff for the summation over the energy levels, $|\varepsilon_{nl}|<\varepsilon_c$, where $\varepsilon_c=\sqrt{\Lambda^2+M_0^2}$. 
While it is hard to be proven analytically, we have checked numerically that for $ \Lambda\rightarrow\infty$, the cutoff dependence on the left-hand and right-hand sides cancels each other and the variational equation
(\ref{2dgap}) gives a cutoff independent result.  Figure \ref{fig1} gives an example for system size $M_0R=10$ at zero temperature and vanishing rotation. It is clear that with increasing cutoff $\Lambda$, the calculation converges, leading to a cutoff independent result.

%%%%%%%%%%%%%%%%%%%%%%%%%%%%%%%%%%%%%
\section{ Results and discussion}\label{s4}
%%%%%%%%%%%%%%%%%%%%%%%%%%%%%%%%%%%%%

Solving the BdG equation simultaneously with the variational equation is not an easy job.  Two convergence issues should be treated carefully. First, since the Bessel function $J_l(x)$
has infinite number of zeros, we need a truncation $j<j_{\rm max}$ to solve the matrix equation (\ref{BdG-Matrix}). This amounts to a high-energy cutoff for the energy levels $\{\varepsilon_{nl}\}$. Second, according to the asymptotic behavior  
 $J_l(x)\sim(2\pi l)^{-1/2}(e x/2 l)^l $ for large $l$, we expect that the summation of $l$ converges fast.  The computational cost depends on the size of the system.  A larger system size $R$ leads to more computational cost.

 %%%%%%%%%%%%%%%%%%%%%%%%%%%%%%%%%%%%%%%%%%%%%%%%%%%%%%%%%%%%%%%%%%%%%%%
\begin{figure}
\centering{}\includegraphics[width=0.45\textwidth]{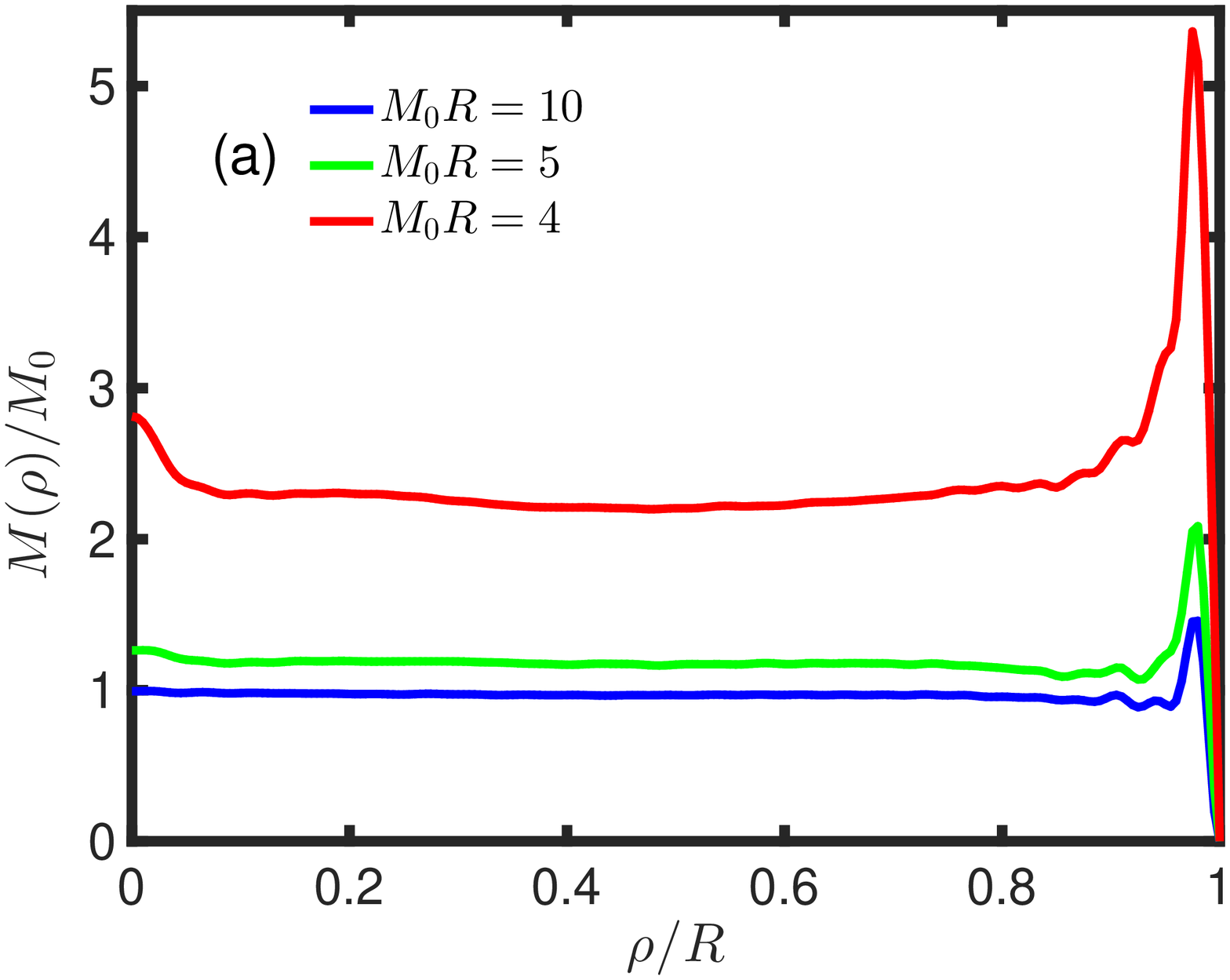} 
\includegraphics[width=0.45\textwidth]{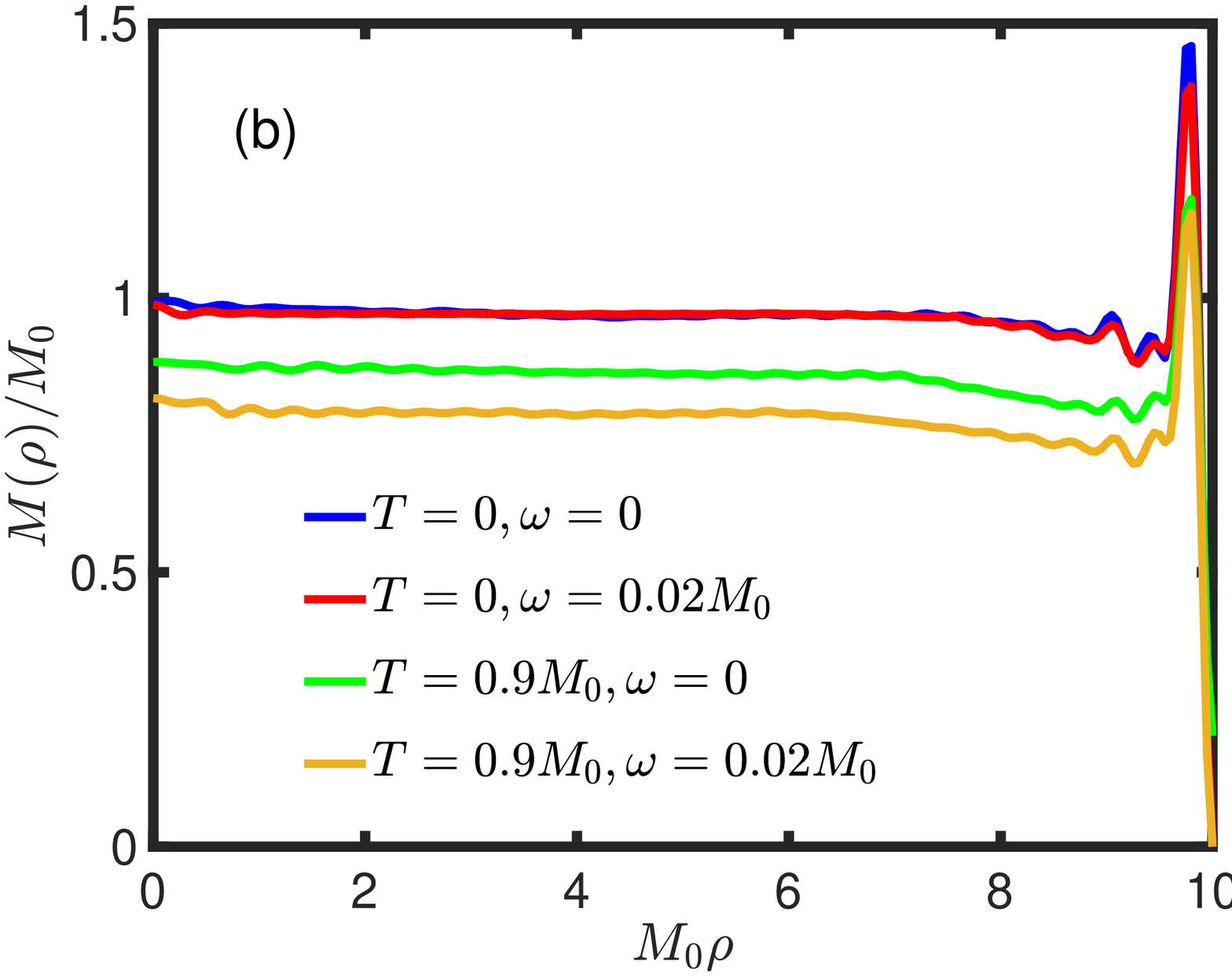} 
\caption{ Results for the trivial solution $\kappa=0$. (a) Finite-size effect: Profile of $M(\rho)$ for various system sizes at $T=0$ and $\omega=0$. (b) Temperature and rotation effects: 
Profile of $M(\rho)$ for different values of $T$ and $\omega$ with a system size $M_0R=10$.
\label{fig2}}
\end{figure}
%%%%%%%%%%%%%%%%%%%%%%%%%%%%%%%%%%%%%%%%%%%%%%%%%%%%%%%%%%%%%%%%%%%%%%%%

We first study the trivial solution with vanishing circulation $\kappa=0$. Fig.~\ref{fig2}(a) shows the pure finite-zise effect at zero temperature and at vanishing rotation. For a large system size $M_0R=10$,  we find that the 
bulk limit is almost reached. The condensate profile $M(\rho)$ is flat in a wide regime and almost reaches the bulk value $M\simeq M_0$. Near the boundary, the condensate ocssilates and finally vanishes at the boundary $\rho=R$. Our self-consistent result thus justifies the LDA for a large system size. However, for small system sizes, our results deviate significantly from the LDA.  Finite-size effect leads to a global enhancement of the chiral condensate and  inhomogeneity around the origin and near the boundary. 

Fig.~\ref{fig2}(b) shows the temperature and rotation effects on the chiral condensate for a large system size $M_0R=10$.  At zero temperature, the rotation does not lead to a suppression of the chiral condensate.
While the profile of the chiral condensate has a slight change due to the finite-size effect,  for an infinite system ($R\rightarrow\infty$), we expect that the rotation has no effect on the chiral condensate, leading to the conclusion that the cold vacuum does not rotate~\cite{NJL-Rotation04,NJL-Rotation05}. While in our BdG formalism it is hard to prove this analytically, we have numerically checked the eigenvalue spectrum $\{\varepsilon_{nl}\}$, with a quantity $(l+1/2)\Omega$ subtracted, almost does not change for a large system size, consistent with the analytical observation~\cite{NJL-Rotation02}. At finite temperature, the rotation generally leads to a global suppression of the chiral condensate.

\begin{figure}
\centering{}\includegraphics[width=0.45\textwidth]{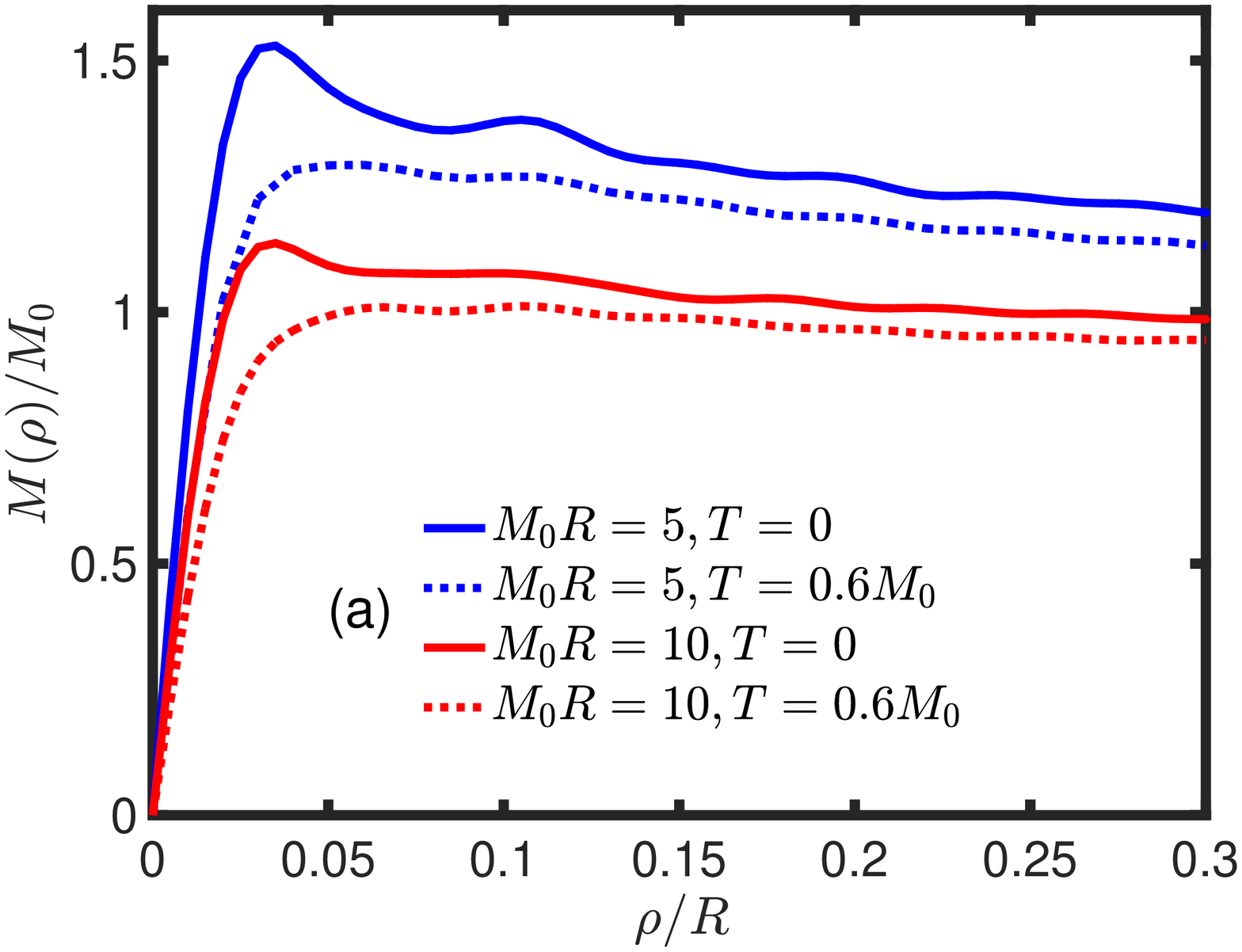} 
\includegraphics[width=0.45\textwidth]{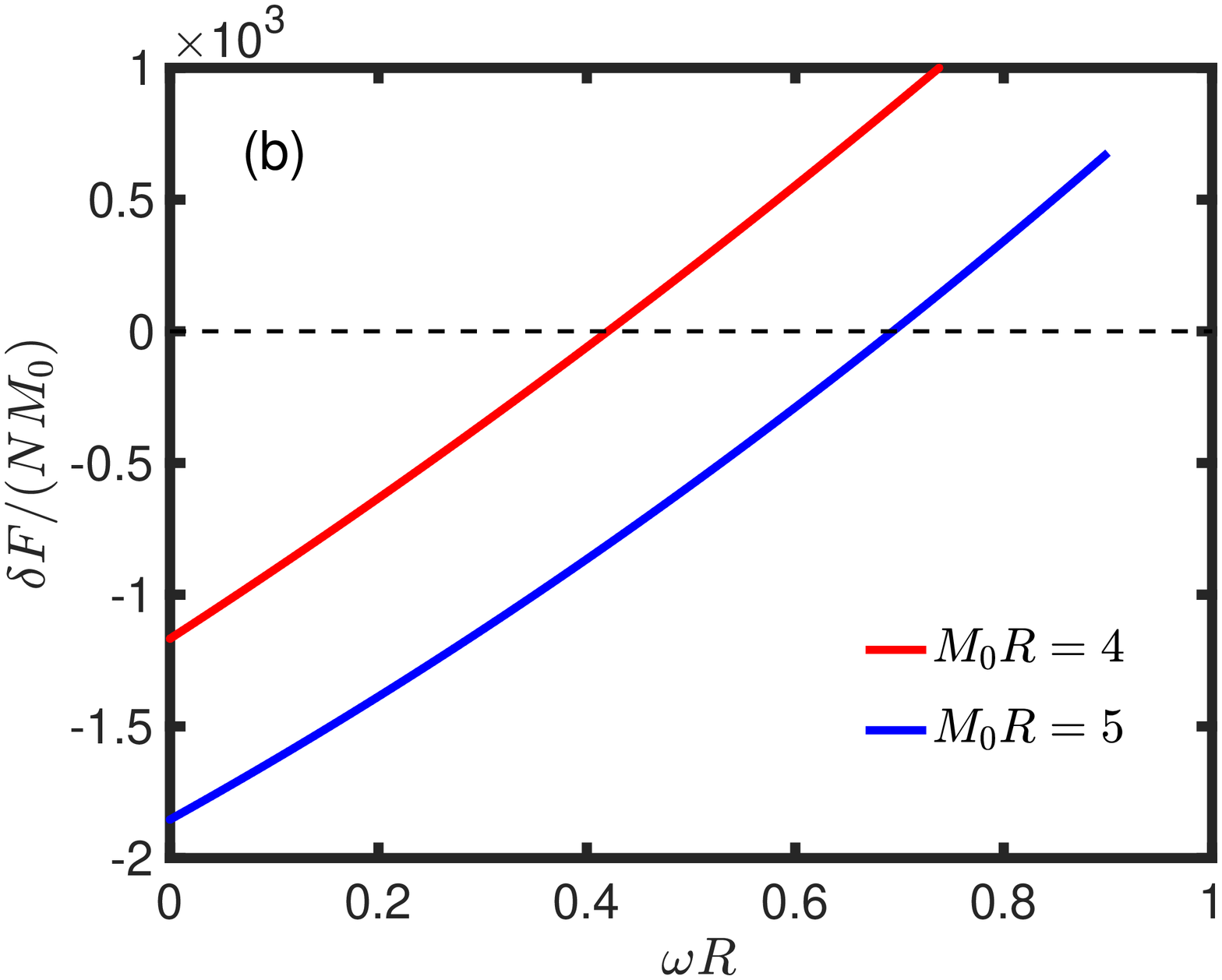} 
\caption{ (a) Vortex core structure with $\kappa=1$: Profile of $M(\rho)$ for different values of $R$ and $T$ at $\omega=0$. (b) Free energy difference $\delta F=F_{\kappa=0}-F_{\kappa=1}$ as a function of $\omega R$
at $T=0$ for two system sizes $M_0R=4$ and $M_0R=5$.
\label{fig3}}
\end{figure}

Next we consider the vortex solution with nonzero circulation $\kappa\neq0$, where the phase of the order parameter, $\phi(\bf r)$, plays a nontrivial role.  Since $\mbox{\boldmath{$\nabla$}}\phi=\kappa\rho^{-1}\hat{\theta}$, the gradient of the phase is singular for $\rho\rightarrow0$,  implying that the kinetic energy associated with the phase would diverge.  The way out is to force the modulus $M(\rho)$ to go to zero for $\rho\rightarrow0$. From the variational equation (16), we obtain $M(\rho)\sim \rho^{|\kappa|}$ for $\rho\rightarrow0$. We note that the vortex solution exists for arbitrary angular velocity $\omega$. Fig.\ref{fig3}(a) shows the typical vortex core structure from our self-consistent BdG calculation for $\kappa=1$ at vanishing rotation. For finite $\omega$, the vortex core structure is qualitatively similar.

%%%%%%%%%%%%%%%%%%%%%%%%%%%%%%%%%%%%%%%%%%%%%%%%%%%%%%%%%%%%%%%%%%%%%%%
%%%%%%%%%%%%%%%%%%%%%%%%%%%%%%%%%%%%%%%%%%%%%%%%%%%%%%%%

For vanishing and slow rotation, the vortex state is an excited state. The excitation energy $E_{v}$ of a vortex for $\omega=0$ can be estimated by the effective Hamiltonian of the phase, $H_{\rm eff}\propto\int d{\bf r}(\mbox{\boldmath{$\nabla$}}\phi)^2$, to the quadratic order. We obtain qualitatively $E_v\propto \kappa^2\ln(R/\xi)$,  where $\xi$ is the size of the vortex core, or the so-called healing length. Since we consider $\omega>0$ without loss of generality, finite rotation reduces the free energy of the vortex state with positive circulation, which can be qualitatively understood by the fact that the vortex carries a finite angular momentum. Therefore, we expect that the vortex state becomes the favorable ground state if the rotation is rapid enough, exceeding a critical angular 
velocity $\omega_c$ which depends on the size $R$, the circulation $\kappa$, and the temperature $T$.

From the estimation $E_v\propto \kappa^2\ln(R/\xi)$, we conclude that the vortex state with circulation $\kappa=1$ would become energetically favorable at large angular velocity. On the other hand, because of the causality constraint $\omega R<1$, the critical angular velocity $\omega_c$ cannot exceed the maximum angular velocity $\omega_{\rm max}=R^{-1}$.  
Fig.\ref{fig3}(b) shows the free energy difference, $\delta F=F_{\kappa=0}-F_{\kappa=1}$, as a function of $\omega R$ for $M_0R=4$ and $M_0R=5$ calculated from our self-consistent BdG approach. It is clear that there exists a critical angular velocity $\omega_c$ below the causality bound. For $\omega>\omega_c$, the vortex state becomes the favorable ground state. In the state, the pseudoscalar condensate is given by $\pi(\rho,\theta)=M(\rho)\sin\theta$, forming an angular stripe-like structure. For large system size, such as $M_0R=10$,  the critical angular velocity exceeds the causality bound and hence becomes unphysical.

The final remark here is on the boundary condition. While we have used a specific boundary condition in this work, the generic feature of the chiral vortices does not reply on it.  We have performed the calculation using a different boundary condition satisfying the current conservation (\ref{current}) and found that the result is not qualitatively changed.  The preference of the vortex state
at large $\omega$ is due to the competition between the vortex excitation energy and the rotation reduction of the free energy, and therefore does not qualitatively rely on the boundary condition. 

%%%%%%%%%%%%%%%%%%%%%%%%%%%%%%%%%%%%%
\section{ Summary }\label{s5}
%%%%%%%%%%%%%%%%%%%%%%%%%%%%%%%%%%%%%

In summary, we have developed a self-consistent approach to study the inhomogeneous chiral condensate and the possible chiral vortices in rotating finite-size matter within four-fermion interacting theories.  
For sufficiently rapid rotation in a finite-size system, the ground state can be a chiral vortex state, a type of topological defect associated with the dynamical chiral symmetry breaking.  The vortex state exhibits angular stripe-like pion condensation, providing a new route to realize pseudoscalar condensation in strongly interacting matter.  In this work, we have studied the simplest scenario, a U$(1)$ chiral symmetry. The extension to SU$(2)$ chiral symmetry may enable us to realize the proposed non-Abelian vortices~\cite{non-Abelian01,non-Abelian02}. Our study on the $2+1$ dimensional system may also have impact on planar condensed matter systems. Since the model can be mapped to a superconducting problem with Dirac fermions, our theoretical predictions may also be tested in some novel superconducting materials, such as graphene~\cite{graphene}.

{\bf Acknowledgment:}    The work  is supported by the National Natural Science Foundation of China, Grant No. 11575093(LW and PZ), 11890712(LW, PZ and LH), 11775123(LH), 11875002(YJ) and the National Key R\&D Program of China(Grant No. 2018YFA0306503)(LH). LW and YJ are also supported by the China Scholarship Council (CSC) for visiting at the University of Tokyo and the Zhuobai Program of Beihang University respectively.

\appendix

\section{BdG formalism for $3+1$ dimensions}
In $3+1$ dimensions, we employ the Weyl representation for gamma matrices. The eigenvalue equation or the BdG equation can be expressed as
\begin{equation}
\left( \begin{array}{cc}
\hat{K}_{11} & \hat{K}_{12} \\ 
\hat{K}_{21}& \hat{K}_{22}
\end{array} \right)
\left(\begin{array}{c}
		u_{n}({\bf r})\\ 
		v_{n}({\bf r}) \end{array}\right)
=\varepsilon_n\left(\begin{array}{c}
		u_{n}({\bf r})\\ 
		v_{n}({\bf r}) \end{array}\right).
\end{equation}
where the blocks of the $\hat{K}$ operator are given by 
\begin{eqnarray}
&&\hat{K}_{11}=i\mbox{\boldmath{$\sigma$}}\cdot\mbox{\boldmath{$\nabla$}}-\omega (\hat{l}_z+\sigma_3/2),\nonumber\\
&&\hat{K}_{22}=-i\mbox{\boldmath{$\sigma$}}\cdot\mbox{\boldmath{$\nabla$}}-\omega (\hat{l}_z+\sigma_3/2),\nonumber\\
&&\hat{K}_{12}=\Delta({\bf r}),\ \ \ \ \ \  \hat{K}_{21}=\Delta^*({\bf r})
\end{eqnarray}
Working in cylindrical coordinates ${\bf r}=(\rho, \theta, z) $, we look for the solution of the form $\Delta({\bf r})=M(\rho)e^{i\kappa \theta}$. Due to the rotational symmetry in the $x-y$ plane and the
translational symmetry along the $z$ direction, the wave functions can be expressed as 
\begin{eqnarray}
&&u_n({\bf r})=e^{ik_zz}\sum_l\frac{e^{il\theta}}{\sqrt{2\pi}}  \left( \begin{array}{c} u^\uparrow_{nl}(\rho) \\u^\downarrow_{nl}(\rho)e^{i\theta} \end{array}\right),\nonumber\\
&&v_n({\bf r})=e^{ik_zz}e^{-i\kappa \theta}\sum_l\frac{e^{il\theta}}{\sqrt{2\pi}}  \left( \begin{array}{c} v^\uparrow_{nl}(\rho) \\v^\downarrow_{nl}(\rho)e^{i\theta} \end{array}\right).
\end{eqnarray}

To impose a boundary condition, we also consider the current conservation which leads to the condition
\begin{eqnarray}
R\int_{-\infty }^{\infty}dz\int_{0}^{2\pi}d\theta\bar{\psi}\gamma^\rho\psi\Big|_{\rho=R}=0.
\end{eqnarray}
In a simple version of the boundary conditions~\cite{NJL-Rotation01}, the wave functions can be expanded as
\begin{eqnarray}
&&u_{nl}^\uparrow(\rho)=\sum_j c_{nj}^\uparrow\phi_{j,l}(\rho),\nonumber\\
&&u_{nl}^\downarrow(\rho)=\sum_j c_{nj}^\downarrow\phi_{j,l+1}(\rho),  \nonumber\\
&&v_{nl}^\uparrow(\rho)=\sum_j d_{nj}^\uparrow\phi_{j,l-\kappa}(\rho), \nonumber\\
&&v_{nl}^\downarrow(\rho)=\sum_j d_{nj}^\downarrow\phi_{j,l+1-\kappa}(\rho),
\end{eqnarray}
where the orthonormal basis $\phi_{j,l}(\rho)$ is given by Eq. (\ref{basis}). For a given $ l $, the BdG equations thus reduces to a matrix form,
\begin{widetext}
\begin{equation}
\sum_{j'}\left( \begin{array}{cccc}
-(K_{\rm L})_{l}^{jj'} & S_{l}^{jj'} & \Delta_{l}^{jj'} & 0 \\ 
S_{l}^{j'j} & (K_{\rm L})_{-l-1}^{jj'} & 0 & \Delta_{l+1}^{jj'} \\ 
\Delta_{l}^{j'j} & 0 & (K_{\rm R})_{-(l-\kappa)}^{jj'} & -S_{l-\kappa}^{jj'} \\ 
0 & \Delta_{l+1}^{j'j} & -S_{l-\kappa}^{j'j} & -(K_{\rm R})_{l+1-\kappa}^{jj'}
\end{array} \right) \left(\begin{array}{c}
c_{nj'}^{\uparrow} \\ 
c_{nj'}^{\downarrow} \\ 
d_{nj'}^{\uparrow} \\ 
d_{nj'}^{\downarrow}
\end{array}  \right)=\varepsilon_{nl}\left(\begin{array}{c}
c_{nj}^{\uparrow} \\ 
c_{nj}^{\downarrow} \\ 
d_{nj}^{\uparrow} \\ 
d_{nj}^{\downarrow}
\end{array}  \right)
\end{equation}
\end{widetext}
where the elements are given by 
\begin{eqnarray}
&&(K_{\rm L})_l^{jj'}=\left[k_z+\omega(l+1/2)\right]\delta_{jj'}, \nonumber\\
&&(K_{\rm R})_l^{jj'}=\left[k_z+\omega(l-1/2)\right]\delta_{jj'},  \nonumber\\
&&S_{l}^{jj'}=i\int d\rho\phi_{j,l}(\rho)\left(l+1+\rho\frac{\partial}{\partial_\rho}\right)\phi_{j',l+1}(\rho), \nonumber\\
&&\Delta_{l}^{j'j} =\int \rho d\rho M(\rho)\phi_{j,l}(\rho)\phi_{j',l-\kappa}(\rho).
\end{eqnarray}
The condensate profile $ \Delta({\bf r}) $ should be self-consistently determined by the variational condition ${\delta F}/{\delta \Delta}=0$, which gives
\begin{equation}
\Delta({\bf r})=G\int \frac{dk_z}{2\pi}\sum_{n,l} v^\dagger_n({\bf r}) u_n^{\phantom{\dag}}({\bf r}) (1-2n_{\rm F}(\varepsilon_{nl})).
\end{equation}

In $3+1$ dimensions, the NJL model is not renormalizable. The summation in the right-hand side  of (A8) is divergent and cannot be removed by a fine tuning of the bare coupling $G$.  We thus need a proper regularization
scheme. For example, we can introduce a smooth cutoff function~\cite{NJL-Rotation01}
\begin{equation}
f(k;\Lambda)=\frac{\sinh(\Lambda/{\delta\Lambda})}{\cosh[\varepsilon(k)/{\delta\Lambda}]+\cosh(\Lambda/{\delta\Lambda})},
\end{equation}
with $ \varepsilon(k)=\sqrt{k_{l,j}^2+k_z^2} $, where $ k_{l,j}=\alpha_{j,l}/R $. This function is suppressed for $ \varepsilon>\Lambda $ and the suppression is smoothened by another parameter $ \delta\Lambda $.

\end{document}